%
\let\useblackboard=\iftrue
%
%
\newfam\black
\input harvmac.tex
\def\Title#1#2{\rightline{#1}
\ifx\answ\bigans\nopagenumbers\pageno0\vskip1in%
\baselineskip 15pt plus 1pt minus 1pt
\else
\def\listrefs{\footatend\vskip 1in\immediate\closeout\rfile\writestoppt
\baselineskip=14pt\centerline{{\bf References}}\bigskip{\frenchspacing%
\parindent=20pt\escapechar=` \input
refs.tmp\vfill\eject}\nonfrenchspacing}
\pageno1\vskip.8in\fi \centerline{\titlefont #2}\vskip .5in}

\ifx\answ\bigans\def\tcbreak#1{}\else\def\tcbreak#1{\cr&{#1}}\fi
\useblackboard
\message{If you do not have msbm (blackboard bold) fonts,}
\message{change the option at the top of the tex file.}
\font\blackboard=msbm10 scaled \magstep1
\font\blackboards=msbm7
\font\blackboardss=msbm5
\textfont\black=\blackboard
\scriptfont\black=\blackboards
\scriptscriptfont\black=\blackboardss

\else

\fi
%
\def\yboxit#1#2{\vbox{\hrule height #1 \hbox{\vrule width #1
\vbox{#2}\vrule width #1 }\hrule height #1 }}
\def\fillbox#1{\hbox to #1{\vbox to #1{\vfil}\hfil}}
\def\ybox{{\lower 1.3pt \yboxit{0.4pt}{\fillbox{8pt}}\hskip-0.2pt}}

\def\comments#1{}

\def\II{\relax{I\kern-.07em I}}

\def\IZ{\relax\ifmmode\mathchoice
{\hbox{\cmss Z\kern-.4em Z}}{\hbox{\cmss Z\kern-.4em Z}}
{\lower.9pt\hbox{\cmsss Z\kern-.4em Z}}
{\lower1.2pt\hbox{\cmsss Z\kern-.4em Z}}\else{\cmss Z\kern-.4em
Z}\fi}
\def\IB{\relax{\rm I\kern-.18em B}}
\def\IC{{\relax\hbox{$\inbar\kern-.3em{\rm C}$}}}
\def\ID{\relax{\rm I\kern-.18em D}}
\def\IE{\relax{\rm I\kern-.18em E}}
\def\IF{\relax{\rm I\kern-.18em F}}
\def\IG{\relax\hbox{$\inbar\kern-.3em{\rm G}$}}
\def\IGa{\relax\hbox{${\rm I}\kern-.18em\Gamma$}}
\def\IH{\relax{\rm I\kern-.18em H}}
\def\II{\relax{\rm I\kern-.18em I}}
\def\IK{\relax{\rm I\kern-.18em K}}
\def\IP{\relax{\rm I\kern-.18em P}}

\font\cmss=cmss10 \font\cmsss=cmss10 at 7pt
\def\IR{\relax{\rm I\kern-.18em R}}

\def\IZ{\relax\ifmmode\mathchoice
{\hbox{\cmss Z\kern-.4em Z}}{\hbox{\cmss Z\kern-.4em Z}}
{\lower.9pt\hbox{\cmsss Z\kern-.4em Z}}
{\lower1.2pt\hbox{\cmsss Z\kern-.4em Z}}\else{\cmss Z\kern-.4em
Z}\fi}
\def\IB{\relax{\rm I\kern-.18em B}}
\def\IC{{\relax\hbox{$\inbar\kern-.3em{\rm C}$}}}
\def\ID{\relax{\rm I\kern-.18em D}}
\def\IE{\relax{\rm I\kern-.18em E}}
\def\IF{\relax{\rm I\kern-.18em F}}
\def\IG{\relax\hbox{$\inbar\kern-.3em{\rm G}$}}
\def\IGa{\relax\hbox{${\rm I}\kern-.18em\Gamma$}}
\def\IH{\relax{\rm I\kern-.18em H}}
\def\II{\relax{\rm I\kern-.18em I}}
\def\IK{\relax{\rm I\kern-.18em K}}
\def\IP{\relax{\rm I\kern-.18em P}}

\font\cmss=cmss10 \font\cmsss=cmss10 at 7pt
\def\IR{\relax{\rm I\kern-.18em R}}

\def\tilde{\widetilde}
\def\frac#1#2{{{#1} \over {#2}}}

\def\prd#1#2#3{{\it Phys. Rev.} {\bf D#1,} #2 (19#3)}

\Title{ \vbox{\baselineskip12pt\hbox{hep-th/9705175}
\hbox{RU-97-37,UTTG-16-97}}}
{\vbox{\centerline{String Dualities from Matrix Theory}}}

\centerline{Micha Berkooz${}^1$ and Moshe Rozali${}^2$}
\smallskip
\smallskip
\centerline{${}^1$ Department of Physics and Astronomy}
\centerline{Rutgers University }
\centerline{Piscataway, NJ 08855-0849}
\smallskip
\smallskip
\centerline{${}^2$ Theory Group,  Department of Physics}
\centerline{University of Texas}
\centerline{Austin, Texas 78712}

\bigskip
\bigskip
\noindent

We suggest that the (2,0) six dimensional field theory compactified on
$S^1\times K3$ is the Matrix model description of both M-theory on $K3$
and the Heterotic string on $T^3$. This proposal is different from
existing proposals for the Heterotic theory. Different limits of the base
space geometry give the different space-time interpretations, making
M-theory/Heterotic duality manifest. We also present partial results
on Heterotic/F-theory duality.

\def\prd#1#2#3{{\it Phys. Rev.} {\bf D#1,} #2 (19#3) }

\Date{May 1997}

\nref\bfss{T. Banks, W. Fischler, S. Shenker and L. Susskind,
hep-th/9610043, \prd{55}{112}{97}.}
\nref\moshe{M. Rozali, hep-th/9702136.}
\nref\brs{M. Berkooz, M. Rozali and N. Seiberg, hep-th/9704089.}
\nref\trieste{P. Horava, Lecture in Trieste Spring School, April 1997.}
\nref\nati{N.Seiberg, hep-th/9705117.}
\nref\tomnati{T. Banks and N. Seiberg, hep-th/9702187;
S. Sethi and L.Susskind hep-th/9702101.}
\nref\various{P. K. Townsend, Phys. Lett. B354:247,1995;
E. Witten, hep-th/9503124, Nucl. Phys. B443:85, 1995.}
\nref\vafaf{C. Vafa, hep-th/9602022, Nucl. Phys. B469:403, 1996.}
\nref\twob{ P. Aspinwall, hep-th/9508154; J. Schwarz,
Phys. Lett. B367:97,1996.  }
\nref\tommotl {T. Banks and L. Motl, hep-th/9703218.}
\nref \petr {P. Horava, hep-th/9705055.}
\nref\evashumit{S. Kachru and E. Silverstein, hep-th/9612162.}
\nref\het{U. Danielsson and G.Ferretti, hep-th/9610082;
L. Motl, hep-th/9612198; 
N. Kim and S.J. Rey, hep-th/9701139.}
\nref\lowe {
 D.A. Lowe, hep-th/9702006;
D.A. Lowe, hep-th/9704041.}
\nref\fhrs { W. Fischler, E. Haylo, A. Rajaraman and L. Susskind,
hep-th/9703102.}
\nref\arvwilly{W. Fischler and A. Rajaraman, hep-th/9704123.}
\nref \enhance {M.R. Douglas, hep-th/9612126.}
\nref\polchb{J. Polchinski, hep-th/9606165, Phys. Rev. D55:6423, 1997.}
\nref\mikegreg{M. Douglas and G. Moore, hep-th/9603167.}
\nref\zeroale{M. Douglas, H. Ooguri and S. Shenker, hep-th/9702203.}
\nref\MikeDon {M.R. Douglas and D. Finnell, to appear.}
\nref \sethi {S. Sethi, private communication.}
\nref\bcd{ D. Berenstein, R. Corrado and J. Distler, unpublished.}
\nref\polch{J. Polchinski, unpublished; M. Dine and Y. Shirman
, Phys. Lett. B377:36,1996. }
\nref\ozhori    {K. Hori and Y. Oz, hep-th/9702173.}
\nref\taylor{W. Taylor, Phys. Lett B394:283,1997.}
\nref\harvstrom{J. Harvey and A, Strominger,
Nucl. Phys. B449:535,1995.} 
\nref\senf{A. Sen. hep-th/9605150, Nucl. Phys. B475:562,1996.}
\nref\vafamorr {C. Vafa and D.R. Morrison, Nucl. Phys. B476:437,1996.}
\nref\witten{E. Witten, hep-th/9610234.}
\nref\schsen{J. Schwarz and A. Sen, hep-th/9507027, Phys. Lett B357:323,1995.}
\nref\wittentwo{E. Witten, hep-th/9507121, Contributed to STRINGS 95:
Future Perspectives in String Theory, Los Angeles, CA, 13-18 Mar
1995.}
\nref\andytwo{A. Strominger, hep-th/9512059, Phys. Lett. B383:44,1996.}
\nref\wittenmoretwo{E. Witten, hep-th/9512219, Nucl. Phys. B463:383,1996.}
\nref\mirror{ A. Strominger, S.-T. Yau and E. Zaslow, hep-th/9606040,
Nucl. Phys. B479:243, 1996.} 
\nref\govin{S. Govindarajan, hep-th/9705113.}

\newsec{Introduction}  

In the last few months a significant amount of evidence has
accumulated in support of the conjecture of Banks, Fischler, Shenker
and Susskind on the non-perturbative formulation of
M-theory\refs{\bfss}. The conjecture, however, needs to be extended
when describing M-theory compactified on a 4-torus down to seven
dimensions. The reason is that the ``Super-Yang-Mills (SYM) on the
dual manifold'' prescription guarantees that the extreme IR of the
base-space will be roughly correct, but does not provide any
information about its UV. The details of the extension were discussed
in \refs{\moshe,\brs}, where it was suggested that M-theory on a
4-torus is given by the large N limit of the (2,0) supersymmetric
field theory compactified on a 5-torus.

As it is misleading to begin with the SYM prescription, our starting
point will be the (2,0) field theory. The  (2,0) theory is
discussed in \refs{\wittentwo-\wittenmoretwo,\nati}. In this paper we discuss
compactifications to seven and eight dimensions that have 8 linearly
realized supersymmetries (in the infinite momentum frame). These are
M-theory on $K3$, the Heterotic string on $T^3$ and $T^2$ and F-theory on
$K3$. We obtain such a theory by compactifying the (2,0) theory on a
5-dimensional base-space that breaks half the supersymmetry. The
manifold which we use is the natural candidate $S^1\times K3$. We will
refer to this manifold as the base-space.

In different degenerations of the $S^1\times K3$ base-space, we obtain
different spacetime interpretations. When the manifold degenerates to
a 4 dimensional base-space in different ways we obtain M-theory on
$K3$ and the Heterotic theory on $T^3$.  As these configurations are
smoothly connected, and we can follow the transition from one limit to
another, the duality between these theories is manifest.  

M-theory on $K3$ is described, in the base-space IR, by a SYM on
another $K3$, with additional degrees of freedom. For the Heterotic
theory on $T^3$ our proposal differs from existing
proposals\refs{\tommotl,\petr}. In this case the base-space $K3$
degenerates in a complicated way, which is well-defined only in some
limits.  We also discuss the origin of the fermions and 8-branes of
\refs{\tommotl-\lowe} in our picture. This is discussed in section 2.

When the manifold degenerates, in a particular way, to a 3 dimensional
manifold we obtain the Heterotic string on $T^2$ and F-theory on
$K3$. For this duality\refs{\vafaf} we obtain only partial results, as
we can obtain F-theory only in a limit of its moduli space. In the
general case we encounter a strongly coupled SYM theory, which limits
our analysis. Again our proposal for the Heterotic string is different
then existing proposals. For F-theory, in a certain limit, our
proposal is an extension of the construction in \refs{\tomnati}
 of the type IIB string. This
is presented in section 3.

Many of the results in this paper were derived independently by
P. Horava, and were reported in \refs{\trieste}.  As we completed this
paper, we received a paper that has some overlap with the results in
section 2 \refs{\govin}.

\newsec{M-Theory on $K3$ and Heterotic strings on $T^3$}
 
Given the (2,0) theory on any manifold, spacetime is obtained only in
certain limits of the base space geometry, when the (2,0) theory can
be approximated by a Kaluza-Klein reduction to a semi-classical SYM
theory \refs{\brs}. In this section we consider the spacetime
interpretations of the (2,0) theory on $S^1\times K3$ in the limits of
the base space geometry in which one of the dimensions is much smaller
than the others. The Kaluza-Klein reduction is then a 4+1 SYM on a
resulting 4-manifold (if such an object exists and is not too
singular), and the space-time that emerges has seven non-compact
dimensions. The resulting theory has 8 linearly realized
supersymmetries which implies that the spacetime theory has a total of 16
supersymmetries. A natural conjecture is that this model defines the
Heterotic string on $T^3$ or M-theory on $K3$. We show below that we
obtain these two spacetime theories as different limits of the
geometry of the base space $S^1\times K3$.

\subsec{M-Theory on $K3$}

We begin by obtaining M-theory on $K3$ in Matrix theory
\refs{\arvwilly}. Let us denote the size of the
$S^1$ by $\Sigma_1$ and the volume of the $K3$ by $V$. The limit of
M-theory on a large $K3$ is obtained by 
\eqn\mlimt{\matrix{\Sigma_1, V\rightarrow 0\cr\cr
{\Sigma_1\over V}\ fixed.}}
The second requirement guarantees that the eleven
dimensional Planck scale is fixed. 

When $\Sigma_1$ is much smaller than any length scale of the $K3$, the
IR of the base-space theory is approximated by a Kaluza-Klein
reduction on the $S^1$, which is a weakly coupled SYM on
$K3$\footnote{$^1$}{It was suggested in
\refs{\petr} that a related 4+1 SYM might have non-zero instanton
number. We do not discuss this issue here
 \refs{\sethi}.} \refs{\bcd} (the gauge
coupling of the this SYM is $g^2=\Sigma_1$). Via T-duality
\refs{\taylor,\ozhori} we then obtain a
description of the moduli space of the theory in terms of 0-brane
moving on the dual $K3$, which is the physical space-time $K3$.

In the case of orbifold limits of $K3$, the requirement that $S^1$ is
smaller than any length scale of $K3$ cannot be satisfied, as there
are 2-cycles of zero size.  This results in the appearance of
additional degrees of freedom in the effective 4+1 SYM. These are
related to the 32 fermions that appear in \refs{\tommotl - \lowe}
and will be discussed further in section 2.1.1. For now we restrict
our attention to degrees of freedom that live in the bulk, far away
from any orbifold points.

For convenience, let us work in the limit
that $K3$ is the orbifold $T^4/Z_2$, whose sizes are
$\Sigma_2,..\Sigma_5$. 
The relation of the spacetime parameters to the SYM parameters are similar to those of M-theory on a
4-torus, and are given in \refs{\moshe,\fhrs}. These are
\eqn\tfour{\matrix { {L_i}^2 = {2 \pi R V \over \Sigma_1 {\Sigma_i}^2}\cr
{L_p}^6= {R^3 V \over (2 \pi)^3 \Sigma_1}}.}
Where $L_i$ (i= 2,...,5) are the spacetime lengths and $L_p$ is the
eleven dimensional Planck length.

In the limit when $g^2 \rightarrow 0$ ($\hbar\rightarrow0$) the
theory becomes semi-classical and the Wilson lines define a moduli
space. This moduli space is interpreted as the classical
compactification manifold in spacetime.  The weakly coupled 4+1 SYM
description of this space is equivalent to the description of 0-branes
moving on this  manifold, as can be shown by a T-duality. This
description is valid when this spacetime manifold is much larger than
$L_p$.

Another check that we have identified correctly the Matrix theory is to
reproduce the moduli space of M theory on $K3$. As is often the case
in the infinite momentum frame, modifications to the ground state of
the theory are obtained by modifying the Hamiltonian. In our
case we can modify the base space geometry. The different choices of
base space geometry should give the spacetime moduli space.

We are therefore interested in 5 dimensional manifolds that break
half the supersymmetry. First there are several discrete choices
\footnote{$^2$}{We are indebted to P. Aspinwall for supplying us
with this information.}. These choices are either $K3\times S^1$, or
$K3\times S^1$ modded out by some discrete group. It is interesting to see
whether the latter choices give new theories in seven dimensions
\footnote{$^3$}{The counting of Moduli seems to be different. See also 
\refs{\schsen}}, but
here we focus on the first option\footnote{$^4$}{To completely define
the theory we need also to choose a spin structure
\refs{\witten}. We are not sure how that affects our discussion.}. 
Once we have made all the discrete choices, the only parameters are
those which define a metric on $S^1\times K3$ with an $SU(2)$
holonomy. The only parameters of such a metric are the size of the
$S^1$ and a choice of an Einstein (Ricci flat) metric on $K3$ (there
can be no components of the metric that mix the $K3$ and the
$S^1$). The Moduli space is therefore locally $SO(3,19)/(SO(3)\times
SO(19))\times R^+$. This is the correct moduli space of M-theory on $K3$
(and of the Heterotic string on $T^3$) \refs{\various}.

The model also has enhanced gauge symmetries at the correct points
of moduli space. If the base-space has certain singularities,
then the T-dual $K3$ has similar singularities\footnote{$^5$}{We are
indebted to S. Kachru for a discussion of this point.}. It was shown
in \refs{\enhance,\arvwilly} that the $N\rightarrow\infty$ Matrix theory then
has the additional states which make up the additional gauge bosons.

We conclude this section in a computation which will be useful
later. To get the correct result after T-dualizing to the zero-brane
picture we need to specify the correct boundary conditions on the gauge
fields. Here we state the boundary conditions which are derived from
this requirement and later we will use them to obtain the boundary
conditions in the Heterotic limit. The boundary conditions on the
gauge fields are \refs{\enhance,\mikegreg,\MikeDon,\sethi}
\eqn\bndry{\matrix{
A_\mu(x)=-\gamma A_\mu({\tilde x})\gamma,\ \mu=1..4\cr\cr
A_0(x)=\gamma A_0({\tilde x}) \gamma \cr\cr Y_i(x)=\gamma Y_i({\tilde
x})\gamma }} where $\gamma$ is some matrix such that $\gamma^2=1$, and
$\tilde x$ is the image of $x$ under the $Z_2$ action. The Matrix
model includes all choices of $\gamma$ \refs{\enhance}. Note that
under these boundary conditions, the instanton
number\footnote{$^6$}{The instanton number in the 4+1 SYM is
identified with momentum in the small circle direction.  } remains
invariant.  This is consistent with the product structure of the base
space - the orbifold group $Z_2$ does not act on the small circle.

We are interested in lifting the boundary conditions to the (2,0)
theory. Since we do not know how to write a theory of $B$ fields in
a $U(N)$ invariant way, lifting the boundary conditions is justified
only for $N=1$. In this case we can take $\gamma=1$. Near an ALE point
in spacetime such a projection on a zero-brane corresponds to a 2
brane wrapped on the shrunken cycle in the ALE
\refs{\enhance,\polchb}. Since $A_\nu=B_{1\nu}$ the boundary conditions
we obtain are
\eqn\sixdbndr{\matrix{
B_{\mu\nu}(x)=\chi(\mu)\chi(\nu)B_{\mu\nu}({\tilde x})\cr\cr
Y^i(x)=Y^i({\tilde x})
}}
where $\chi(\mu)=-1$ for $\mu=2,3,4,5$ and $\chi(\mu)=1$ for
$\mu=0,1$. We will return to these boundary conditions later.

\medskip
\noindent{\it 2.1.1 Additional Degrees of Freedom}
\medskip

As we have seen in the Matrix description of M-theory on $T^4$
\refs{\brs}, the process of obtaining the world-volume description
from a spacetime picture is incomplete. In general the spacetime
description probes only the IR of the base space theory, and in order
to define it one has to add UV information. In the present case, the
spacetime-based prescription of ``T-dualizing 0-branes on a $K3$'' fails
to capture the complete set of IR degrees of freedom if not done
carefully \refs{\MikeDon,\sethi}.

To demonstrate this, let us discuss the (2,0) theory along its flat
directions, where we know how to write it down. In the $T^4/Z_2$ limit
of the $K3$ we have 16 shrunken 2-cycles and another 10 finite
2-cycles that are associated with the bulk of $S^1 \times T^4/Z_2$. The
10 cycles in the bulk become the gauge field after we reduce on
$\Sigma_1$. The other 16 cycles, however, also contribute massless
degrees of freedom. These are equal to $\int B$ on the shrunken
2-cycles. In fact, these are the 16 chiral bosons of the Heterotic
string in \refs{\harvstrom}, and will be identified with the 32
fermions in the Matrix picture of the Heterotic string on $T^3$ which
we discuss shortly. Whether they will remain massless when we take
$\Sigma_1\rightarrow 0$ depends on their periodicity along the $S^1$
direction. In the Heterotic case \refs{\tommotl} half the fermions are
periodic along the $S^1$ and half are anti-periodic. Borrowing this
result for our case (we did not derive it from the (2,0) theory)
suggests that half of them remain massless in this limit.

On the other hand, the analysis of 0-branes moving on the physical
$T^4/Z_2$ does not require these additional degress of freedom, and
T-dualizing this picture  gives only the Wilson lines in the
bulk.  

There are additional degrees of freedom that the 0-branes might
miss. Suppose we slightly blow up the orbifold. This is given
\refs{\mikegreg,\zeroale} by a small deformation of the 0-brane
Hamiltonian.  The masses of strings that stretch between 0-branes are
likewise slightly perturbed. In our suggestion the
situation is different. Here we have also slightly blown up certain
singularities and there are now configurations of the B-field in which
it varies on the small cycle. These are new massive degrees of freedom
whose mass scale is inversely proportional to the size of the small
cycles. It would seem that the local picture of 0-branes near a
slightly blown-up ALE point misses these states, which exist in
our proposal.

\subsec{Heterotic Theory on $T^3$}

\noindent{\it 2.2.1  Heterotic vacuum with $SU(2)^{16}$ Enhanced 
Gauge Symmetry}
\medskip

The case in which $K3=T^4/Z^2$ (which has 16 $A_1$ singularities) is
the easiest to analyze. This configuration is the one that is most
closely related to the configuration of \refs{\tommotl,\petr}. Let us
pick one dimension of $T^4$, say $\Sigma_5$, and take it to zero, as
well as the volume of the remaining space
$V=\Sigma_1\Sigma_2\Sigma_3\Sigma_4$.  Again, we take $V$ and
$\Sigma_5$ to zero at a fixed ratio.

After the Kaluza-Klein reduction on $\Sigma_5$ we obtain a SYM on $S^1\times
(T^3/Z_2)$ with a coupling $g^2=\Sigma_5$ and some boundary conditions
on the gauge fields. These are obtained from the boundary conditions
on the $B$ fields. For N=1 we can check this explicitly. The
Kaluza-Klein reduction is obtained by defining $A_\mu=B_{5\mu},\
\mu=1,2,3,4$ and using \sixdbndr\ (for N=1). The boundary conditions
 one obtains are
\eqn\hetsybndry{\matrix{
A_{0,1}(\sigma,\sigma^i)=-A_{0,1}(\sigma,-\sigma^i)\cr\cr
A_a(\sigma,\sigma^i)=A_a(\sigma,-\sigma^i),\ a=2,3,4\cr\cr
Y^i(\sigma,\sigma^i)=Y^i(\sigma,-\sigma^i)
}}
which are the same  as in \refs{\tommotl}. Note also that the instanton
number reverses its sign under the $Z_2$ action, which means that 
these boundary conditions give
 the unique correct extension to  the 5-dimensional
manifold $S^1 \times T^4/Z_2$.
\footnote{$^7$}{As was explained in \refs{\enhance}, the state
with N=1 corresponds to a solitonic state in the M-theory. We are now
in position to see how the gauge boson changes from a perturbative
state in the Heterotic string to a solitonic state in M-theory.}

We have obtained an Heterotic string theory on $T^3$, and we can write its
parameters in terms of $g^2,\Sigma_{1,2,3,4}$. It is more instructive,
however, to write it in terms of the M-theory on $K3$ parameters \tfour. Doing so,
one obtains
\eqn\hetmdual{\matrix{
T_{string}={L_2L_3L_4L_5\over (2\pi)^5L_p^6}\cr
\lambda_7^4={(L_2L_3L_4L_5)^3\over (2\pi)^2L_p^{12}}
}}
which are the seven dimensional Heterotic/M-theory duality
relations \refs{\various}. One can also reproduce more detailed formulas
that relate the radii of the $K3$ to those of the $T^3$\refs{\polch}.

\medskip
\noindent{\it 2.2.2. A Conjecture Regarding the $E_8\times E_8$ Case}
\medskip

We are interested in the Heterotic string on $T^3$ in its M-theory
limit. {\it i.e.} when it is M-theory on $S^1/Z_2\times T^3$. We therefore
expect to see a well defined moduli space of this form only when the space-time gauge symmetry is $E_8\times E_8$, or a
subgroup of it. We are interested not only in the moduli space, but
also in the masses of some of the modes when we go along the flat
directions. These are important in order to reproduce graviton
scattering. 

The picture that we suggest is very similar to that of Vafa and
Morrison \refs{\vafamorr}. Let us check the case in which the
base-space $K3$ has two $E_8$ singularities. The $K3$ can then be written
as an elliptic fibration over  $P^1$. On the $P^1$ there are two
singular fibers which contain the $E_8$ singularities and four
additional singularities (where the fiber but not the K3
degenerate). We are interested in the limit in which a pair of the
additional singularities approach each $E_8$ locus. In that case the
base becomes a long thin cylinder capped in the vicinity of the $E_8$
singularity. Throughout the cylinder, as long as we are away from the
singularities, the fibre has a constant complex structure
parameter. 

 We are interested in reducing the (2,0) theory on the small circle
which is a part of the cylindrical base of the elliptic fibration. We
take therefore all the other dimensions of the base space to be of the
same order of magnitude, and larger than this small circle. Note that
the size of the fibre is a parameter of the theory (unlike in
F-theory). In this configuration the (2,0) theory has a mass gap and
we can perform a Kaluza-Klein reduction on the small circle.  The base
space of the resulting 4+1 SYM looks like $S^1\times T^2\times I$
where the first $S^1$ is outside the K3, the $T^2$ is the fibre and
$I$ is an interval, which is what is left from the cylinder after the
Kaluza-Klein reduction.  On this space we have a weakly coupled gauge
theory with some boundary conditions on the gauge fields and matter
fields. We can now have four Wilson lines on this space which give us
the four compact space coordinates.

We do not know how to calculate the boundary conditions in this
picture, so we can not verify it. Let us note, however, that the base
space of this SYM, $S^1\times T^2\times I$, is metrically flat. This
means that the higher momentum modes and the energy of W's, which we
integrate out when we calculate graviton scattering, are spaced at
constant intervals in a way that is similar to the spacing of the
stretched strings in the 0-brane picture. It is possible therefore that
it reproduces correctly the graviton scattering amplitude. 

It is interesting to compare this picture to the current
suggestion of the Heterotic string on $T^3$ \refs{\tommotl,\petr}. In that
suggestion the base space is always $S^1\times (T^3/Z_2)$ (perhaps with a
non-trivial metric \refs{\petr}) and the fermions are
not associated with the singularities but rather can move about. The
fundamental domain of the $Z_2$ action can be taken to be $S^1\times
T^2\times I$ which is the same base space that we obtained. From the
point of view of the base space the only difference is the resolution
of the singularities at the boundaries of the fundamental domain. 

Away from the $E_8\times E_8$ loci, we obtain the full $K3$. For
a general shape of the $K3$ our suggestion differs 
from \refs{\tommotl,\petr}. In general there is no well
defined way to perform a Kaluza-Klein reduction and obtain
a SYM on some 4-manifold. Even if we are able, at certain
limits, to pick a small circle in the $K3$ and reduce on it,
the resulting 4-manifold would be highly singular. It is not
clear in that case how a SYM can be defined on such a manifold.

Another important difference is the way that the fermions are
treated. In our picture the fermions are to be understood as
fermionization of the bosons $\int B$ over shrunken cycles. As such
they are localized at the singularities and are not allowed to
move. Enhanced symmetry is obtained in a geometric way in which the
mixing of the compact space parameters and the $E_8\times E_8$ Wilson
lines is apparent.

More important is the fact that we do not need to add these fermions
by hand \refs{\evashumit}. They are automatically provided by the
(2,0) definition of the theory. At no point of the discussion do we
 need to take a circle, in the Matrix description of M-theory on $T^4$,
orbifold it and add 8-branes. Rather the 8-branes are generated in the
effective space-time by the existence of additional degrees of freedom
in a specific degeneration of the (2,0) base-space.  We know that
there are 8-branes only through the existence of the 32 fermions. When
we calculate any low energy scattering the fermions contribute to the
scattering amplitude such that a low energy observer interprets the
result as the existence of 8-branes.

When we blow up a singularity, to break the space-time gauge symmetry,
we still have the 32 fermions, but now we have many more degrees of
freedom. There are finite energy configurations of the $B$
fields in which they vary on the cycle which  we have blown up. This is again a
departure from the current suggestion for the Matrix description of
the Heterotic string, and it demonstrates again the problems that arise
when we take the space-time picture and try to extrapolate it to the
base-space.

\subsec{Duality}

To summarize, both M/$K3$ and Het./$T^3$ are described by the same
model. In one limit of the geometry of the base-space we obtain the a
description of the weakly coupled low-energy of the Heterotic string
and in another limit that of M theory on $K3$. The transition between
these two limits, as expected, goes through a region in which
the compact part of space-time is not well defined.

\newsec{F-theory on $K3$ and Heterotic on $T^2$ }

Since we have found Heterotic$(T^3)$-M$(K3)$ duality as a simple
geometric relation in the (2,0) field theory, it is interesting to ask
whether we can do the same for the Heterotic/F-theory duality in eight
dimensions \refs{\vafaf}. The Matrix model of the Heterotic theory is
similar to the one we presented in the previous section, and we
present here only a partial analysis of the IIB side. More precisely,
in the region of parameters where we can analyze the base-space
theory reliably, we obtain F-theory compactification on $K3 \times
S^1$ down to seven dimension. Since the radius of the circle will be
much smaller then that of the F-theoretic base of the $K3$, we never
reach the 8 dimensional F-theory. We comment below on possible ways of
analyzing the 8 dimensional limit.

A point of notation. In places that the notation might be ambiguous we
 denote by a subscript F quantities from F-theory. For example $K3_F$
 denotes the elliptically fibered $K3$ that defines the F-theory
 compactification. Similarly a subscript M denotes M-theory
 quantities, and a subscript B denotes quantities that relate to the
 base-space on which we compactify the (2,0) theory.

\subsec{The Heterotic Theory}
  
We begin with the Heterotic theory on $T^2$. We take $K3_B$ to be
elliptically fibered. Let $\Sigma_{2,3}$ denote the sizes of the fibre
and $\Sigma_{4,5}$ the sizes of the base of the elliptic fibration. We
suggest that the Heterotic spacetime will be obtained in the limit in
which the base goes to zero.

For the purposes of scaling we  concentrate on the case of
$K3=T^4/Z_2$. We  work in the following 
limit of the base space geometry 
\eqn\hetlim{\matrix{
\Sigma_{4,5}\rightarrow 0\cr\cr
{\Sigma_5\over\Sigma_4}\ll 1,\ fixed\cr\cr
\Sigma_{1,2,3}\ fixed
}.}
Because of this hierarchy we first
reduce the (2,0) theory on $\Sigma_5$. The result is a SYM theory in
4+1 dimensions with a  coupling $g^2=\Sigma_5$. We then reduce on the
circle of length $\Sigma_4$. In this limit we obtain  3+1 SYM with a
coupling $g^2={\Sigma_5\over\Sigma_4}$ which is weak. The
moduli space of this SYM theory describes the resulting spacetime.
The spacetime dimensions are given in terms of the SYM quantities as
follows:
\eqn\hetscal{\matrix{
{l_k}^2 = {2 \pi R \Sigma _1 \Sigma_2 \Sigma_3 \Sigma_4\over  \Sigma_5
{\Sigma_k}^2}, \ \ k= 1,2,3 \cr\cr
{l_4 }^2 = {2 \pi R \Sigma_1 \Sigma_2 \Sigma_3\over\Sigma_4 \Sigma_5}\cr\cr
{l_{p,11}}^6= {R^3 \Sigma_1 \Sigma_2 \Sigma_3 \Sigma_4\over (2 \pi)^3
\Sigma_5}
}}
Where $l_{p,11}$ is the eleven dimensional Planck length, $l_4$ is a
dimension that decompactifies in the limit of a vanishing base. 

The resulting spacetime is therefore 8 dimensional and describes the
M-theory compactified on $S^1/Z_2\times T^2$. The relevant quantities
that measure the decompactification of the 8th dimension are
$l_4/l_{p,11}$ or $l_4/l_{p,8}$ where $l_{p,8}$ is the eight
dimensional Planck scale. Using \hetscal\ we obtain
\eqn\decomhet{{l_4^2\over
l_{p,8}^2}={(\Sigma_1\Sigma_2\Sigma_3)^{2\over3}\over
\Sigma_4\Sigma_5}.} Therefore, in the  limit \hetlim
   we indeed approach an eight dimensional spacetime.

As expected, $\Sigma_1$, the base space circle, is inversely
proportional to the spacetime orbifold length, $l_1$. A large base
space circle corresponds therefore to a weak Heterotic coupling. The
spacetime 2-torus is dual to the base space torus, which is the
elliptic fiber of the $K3_B$ surface.

As before, the manifold which we obtain after we shrink part of $K3_B$
can be very singular. It is not clear how to define a SYM theory on
it, other than as  the  Kaluza-Klein reduction above. This was an issue
that we have already seen in the case of the Heterotic 
theory on $T^3$, and 
the discussion there
applies here as well.

\subsec{F-theory}

The Heterotic string in eight dimension has a dual description
in terms of F-theory compactified on $K3_F$. The base of $K3_F$
is a compactification manifold for type IIB theory, and the 
fibration describes $SL(2,Z)$ monodromies on the base. The
Heterotic coupling appears in this context roughly as the  
 area of the base (the more precise statement is given below). 
 
Let us first discuss what is our goal. We will be able to attain it
only partially. We are interested in a Matrix description of a certain
vacuum of type IIB string. In \refs{\tomnati}\ the IIB string theory
was constructed in terms of an interacting fixed point field theory in
2+1 dimensions \refs{\nati}. If we want 
to obtain the IIB string on a circle, 
we expect to have (probably only in the IR) a family of 2+1 fixed
point theories that have a prefered scalar. As we change the vev of
the scalar we sweep out the additional circle.

The way to obtain such a picture is to start with M-theory on $T^3$
and take two of the circles of the dual torus to infinity. We can then
obtain this picture by doing a Kaluza-Klein reduction on the small
circle. The preferred scalar is then the Wilson line along this small circle. Another way of rephrasing this construction is the
following. SYM on the 3-torus is analogous to a IIB 3-brane wrapping a
3-torus. When one of the circles of the two-torus goes to zero the IR
is better described by T-dualizing this smaller circle to a IIA
2-brane wrapping the two large circles.

The limit which we now discuss is the following
\eqn\flimit{\matrix{\Sigma_1,\Sigma_4,\Sigma_5\rightarrow 0\cr\cr
\Sigma_1\ll\Sigma_4,\Sigma_5\cr\cr
{{\Sigma_1\over\Sigma_4\Sigma_5}\rightarrow\infty}\cr\cr
\Sigma_2\ll\Sigma_3\ both\ held\ fixed}}

(the reason for taking this limit is to obtain a regime in which
F-theory is described reasonably well by 10D supergravity, as we will
see later). 

 Given this hierarchy, we first reduce the (2,0) theory on $\Sigma_1$,
obtaining a SYM theory in 4+1 dimensions, with a coupling
$g^2=\Sigma_1$, compactified on a $K3_B$ surface. As discussed above
the spacetime interpretation of this theory is M-theory on a $K3_M$.  In
the limit \flimit\ we get a particular degeneration of $K3_M$, in which
the fibre is much smaller than the base. This is the immediate
generalization of the description above for the M-theory on a 3-torus
giving IIB on a circle.  In our case, this is the appropriate
degeneration that yields the F-theory description \refs{\vafaf}.

To analyze the moduli space, it is again convenient to use an analogue
model, which is the following. SYM on a $K3_B$ manifold has an
analogue model which is a IIA 4-brane wrapping the $K3$. We are
interested in a description in which we have a 2-brane wrapping the
fibre. We can accomplish this by performing a T-duality transformation
that takes the volume of the $K3$ to its inverse, and the wrapped
4-brane to a 0-brane \refs{\ozhori}. We then do T-duality on each
fibre, or go to the mirror $K3$ \refs{\mirror}, to obtain a description in terms of
2-branes wrapping the fibre. Out of these three configurations
(wrapped 4-brane, 0-branes or wrapped 2-brane) we should take the one
that describes the IR behavior more accurately. Since we started with
a large fibre and a small base, the wrapped 2-brane has a large base
and a large fibre and is the correct description. We obtained a
picture of an elliptically fibered $K3_F$, with a 2+1 SYM wrapping the
fibres. This is exactly what we would expect as world volume IR
description of the IIB string. This procedure is essentially going to
the IIB  limit by shrinking the fibre in M-theory on $K3$.

Two points are important. One is that this picture only captures the
IR, which is sufficient for our purposes. The second is that we did
our computation in weak coupling (of the SYM of $K3_B$), which means
that it is reliable. This was done by taking $\Sigma_1$ to be much
smaller than any other scale. The down side, as we will see shortly,
is that the base-space of F-theory is much larger than the additional
dimension which grows (when we shrink the two dimensions in M-theory)
so we are not in the  eight dimensional F-theory limit. However, both
the new dimension which grows and the base of $K3_F$ are much larger
than the 10 dimensional IIB Planck  length, so it is a regime in which we
require to have a well defined moduli space.

One can be much more precise about the parameters of the resulting
F-theory (we will work in the orbifold limit). The M-theory on $K3$
parameters were obtained above, in \tfour. In the limit we are
considering in the present context, the $K3_M$ fibres go to
zero. Whenever a spacetime torus of lengths $L_2, L_3$, shrinks in
M-theory, we get type IIB theory with a large circle of length
\refs{\twob}
\eqn\fa{
L_F = { (2 \pi)^3 {L_p}^3 \over L_2
L_3}={\sqrt{2\pi R \Sigma_1\Sigma_2\Sigma_3\over\Sigma_4\Sigma_5}
\rightarrow\infty}}
And the size of the base space of $K3_F$ is 
\eqn\fb{
 L_4 L_5 = { 2 \pi R \Sigma_2 \Sigma_3\over\Sigma_1}\rightarrow\infty
.}
so both the base-space and the additional circle go to infinity.
In particular they grow relative to IIB 10
dimensional Planck scale $l^4_{p,10}\propto R^2\Sigma_2\Sigma_3$ so we
expect to have a good moduli space.

The coupling constant of the
type IIB theory is the complex structure of the base space fiber,
$\Sigma_2 \over \Sigma_3$. More generally the shape of the torus
on which the 2+1 SYM lives changes as we go around the base
space. This is expected in view of  \refs{\tomnati}.

Another quantity which we can calculate is the size of the base
relative to the size of the additional circle 
\eqn\prob{
{ {L_F}^2 \over L_4 L_5} = {{\Sigma_1^2 \over \Sigma_4 \Sigma_5}\ll 1}
}
The additional circle is much smaller than the base, and the theory
does not approach an eight dimensional
theory. However, again,
since both these sizes are larger than $l_{p,10}$ we expect to 
see them in the   moduli space of the base-space theory.
 
The eight dimensional limit can be reached if we take the circle
$\Sigma_1$ to be larger than the $K3_B$ base, unlike the limit in
\flimit. We do not know how to get a moduli space which corresponds to
the F-theory base space in that case.  Reduction on the $\Sigma_1$
first gives natural variables by which to describe the base  of
$K3_F$, but this is justifiable only when $\Sigma_1$ is the smallest
length scale.  In the eight dimensional limit we need to reduce on
$\Sigma_{1,4,5}$. For a generic $K3_B$ this yields 
 a very singular object, but in simple cases (orbifolds for example)
we  obtain a 2+1 dimensional strongly coupled SYM theory. It may be that
there is some D=3 duality by which one will be able to obtain the IIB
space-time.

To summarize we see the base of elliptic fibration of the 
IIB string in the F-theory on
$K3\times S^1$ only in a specific regime. In that case it can be seen
using the full 5+1 dimensional definition of the theory.

\centerline{\bf Acknowledgments}

We would like to thank O. Aharony, P. Aspinwall, T. Banks,
D. Berenstein, R. Corrado, J. Distler, M. Douglas, D. Finnell,
W. Fischler, O. Ganor, P. Horava, S. Kachru, N. Seiberg, S. Sethi and
S. Shenker for useful and illuminating discussions. This work was
supported by DOE grant DE-FG02-96EF40559, by the Robert A. Welch
Foundation and by NSF grant PHY 9511632.

\listrefs
\end